\documentclass[]{spie}  

\usepackage{amsmath,amsfonts,amssymb}
\usepackage{graphicx}
\usepackage[colorlinks=true, allcolors=blue]{hyperref}
\usepackage[ruled,vlined]{algorithm2e}

\newcommand{\yp}{y^i_p}
\newcommand{\yf}{y^i_f}
\newcommand{\up}{u^i_p}
\newcommand{\uf}{u^i_f}

\title{Advanced wavefront sensing and control demonstration with MagAO-X.}

\author[a]{Sebastiaan Y. Haffert}
\author[a]{Jared R. Males}
\author[a]{Kyle Van Gorkom}
\author[a]{Laird M. Close}
\author[a]{Joseph D. Long}
\author[a,b]{Alexander D. Hedglen}
\author[c]{Kyohoon Ahn}
\author[a,b,c,d]{Olivier Guyon}
\author[e]{Lauren Schatz}
\author[a,b]{Maggie Kautz}
\author[a,b]{Jennifer Lumbres}
\author[a,b]{Alexander Rodack}
\author[a,b]{Justin M. Knight}
\author[f]{He Sun}
\author[g,h]{Kevin Fogarty}
\author[e]{Kelsey Miller}

\affil[a]{Steward Observatory, University of Arizona, Tucson, Arizona, United States}
\affil[b]{Wyant College of Optical Science, University of Arizona, 1630 E University Blvd, Tucson, AZ 85719, USA}
\affil[c]{National Astronomical Observatory of Japan, Subaru Telescope, National Institutes of Natural Sciences, Hilo, HI 96720, USA}
\affil[d]{Astrobiology Center, National Institutes of Natural Sciences, 2-21-1 Osawa, Mitaka, Tokyo, JAPAN}
\affil[e]{Kirtland Air Force Base, Air Force Research Laboratory, Albuquerque, NM, USA}
\affil[f]{Department of Computing and Mathematical Science, California Institute of Technology, Pasadena, CA 91125, USA}
\affil[g]{The Division of Physics, Mathematics and Astronomy, California Institute of Technology, Pasadena, CA 91125, USA}
\affil[h]{NASA Ames Research Center, Moffett Field, California 94035 USA}

\authorinfo{Further author information: (Send correspondence to S.Y.H.)\\S.Y.H.: E-mail: shaffert@arizona.edu\\ S.Y.H. is a NASA Hubble fellow}

\begin{document} 
\maketitle

\begin{abstract}
The search for exoplanets is pushing adaptive optics systems on ground-based telescopes to their limits. Currently, we are limited by two sources of noise: the temporal control error and non-common path aberrations. First, the temporal control error of the AO system leads to a strong residual halo. This halo can be reduced by applying predictive control. We will show and described the performance of predictive control with the 2K BMC DM in MagAO-X. After reducing the temporal control error, we can target non-common path wavefront aberrations. During the past year, we have developed a new model-free focal-plane wavefront control technique that can reach deep contrast ($<$1e-7 at 5 $\lambda$/D) on MagAO-X. We will describe the performance and discuss the on-sky implementation details and how this will push MagAO-X towards imaging planets in reflected light. The new data-driven predictive controller and the focal plane wavefront controller will be tested on-sky in April 2022.
\end{abstract}


\section{INTRODUCTION}
One of the primary goals of the upcoming generation of Giant Segmented Mirror Telescopes (GSMT) will be the direct imaging of Earth-like exoplanets. This requires large apertures, both for their light collecting area and their resolution. Planets, like Earth, are usually many orders of magnitude fainter than their host star \cite{traub2010exoplanets}. This makes them difficult to detect. Especially when the planets are only separated by several times the diffraction limit. Furthermore, atmospheric turbulence create large wavefront errors that need to be corrected. High-contrast imaging instruments tackle both problems by removing the influence of the star with extreme adaptive optics and coronagraphs \cite{guyon2018extreme}. Earth-like planets around M dwarf stars, which are the primary targets for the upcoming telescopes, have a contrast that is close to $10^{-8}$\cite{guyon2018wavefront}.

Current direct imaging instruments routinely reach post-processed contrast levels of $10^{-4}$ to $10^{-6}$ at angular separations between 0.1 arcsec and 1.0 arcsec \cite{beuzit2019sphere}. This sensitivity is enough to image and characterize massive self-luminous planets \cite{marley2007hotjupiters} that emit the majority of their emission in the near-infrared part of the spectrum. And even though these instruments are sensitive enough to detect Jupiter-like planet, very few are discovered. Analysis of direct imaging surveys and radial velocity surveys hint that there is a turn over where the occurrence rate of exoplanets starts to drop \cite{bowler2015gpoccurence, nielsen2019gpies, fernandes2019occurence, wagner2019wideoccurence}. This turnover happens between 1 to 10 AU, which is the expected position of the snow line. 

The sensitivity closer in to the star has to be improved for both current exoplanet planet science and future exoplanet science. The limitation at small angular separations for the upcoming GSMTs can be divided in 3 problems:
\begin{itemize}
  \item Time lag in the adaptive optics (AO) system \cite{kasper2012hci,milli2017sphereperformance,cantalloube2019winddrivenhalo}. The correction of the atmosphere is always trying to catch up because the wavefront that has been measured has also already passed through the system. This causes a delay that can not be corrected anymore. The servo-lag error creates a so called wind-driven halo that limits the contrast.

  \item The optics of the coronagraph are not part of the wavefront sensor. This means that light will travel through non-common optical paths, which will create internal instrument aberrations that are not visible to the wavefront sensor. These non-common path aberrations create speckles that leak through the coronagraph.
  
  \item The GSMTs will be segmented. The segmentation create differential piston modes, which are difficult to sense with conventional wavefront sensors. These segment piston modes for the Giant Magellan Telescope or petal modes for the European Extremely Large Telescope are very low-order modes. And, strong low-order modes create the most coronagraphic leakage.
\end{itemize}

In this proceeding we summarize the work that has been done with the MagAO-X instrument to tackle these problems. MagAO-X is a new high-contrast imaging instrument for the 6.5 Magellan Clay Telescope\cite{males2022magaox}. Section 2 describes our approach to predictive control to reduce the wind-driven halo. A new focal-plane wavefront control strategy is shown in Section 3. And Section 4 shows a new approach to measuring and controlling differential piston / petal modes.

\section{Predictive control}
There are multiple approaches to solve the wind-driven halo. The first is to run the AO system at a high enough speeds that the atmospheric turbulence is frozen. This requires measuring the wavefront at speeds of several kHz. This approach use at SCeXAO and MagAO-X \cite{jovanovic2015scexao, males2020magao}. While running faster can reduce the impact of the wind-driven halo it does not solve it completely. The system is still running behind even at high speeds. Another approach is to predict how the atmosphere is going to evolve and correct the wavefront errors before they are measured. Predictive control can lead to significant gains in post-processed contrast for high-contrast imaging. The post-processed contrast could be improved by a factor 100 to 1000, if predictive control is used and the temporal evolution of the atmosphere is predictable \cite{guyon2017eof,males2018lpc,correia2020hcipwfs}. 

We have recently developed the data-driven subspace predictive control (DDSPC) algorithm \cite{haffert2021data}. This algorithm only uses the wavefront measurements and the past DM commands to determine the new optimal command. The DDSPC algorithm directly uses the closed-loop residuals, without reconstructing the full turbulence. The advantage of this approach is that we side step any reconstruction error due to model errors. This approach was implemented for the GPU with the Compute Unified Device Architecture (CUDA) \cite{cuda} to create a real time adaptive controller. The controller runs at 2kHz in double precision and at 4 kHz in single precision for MagAO-X, which has 1600 controlled modes.

The DDSPC algorithm is called data-driven and model free. This is a slight misnomer, because there is no controller that is truly model free. If models are called model free then what is meant is that there is no underlying parametric model of which the parameters are optimized. The DDSPC algorithm uses an auto-regressive structure as its backbone. This results in the following model structure for a single DM mode or actuator\cite{haffert2021data},
\begin{equation}
    \yf = A \yp + B \up + C \uf.
    \label{eq:prediction}
\end{equation}
Here $\yf$ and $\uf$ are the future vectors that contains the future $N$ wavefront sensor measurements and DM commands at time step $i$. And $\yp$ and $\up$ are the vectors that contain the $M$ past wavefront sensor measurements and DM commands at time step $i$. This is a model that is completely linear in $A$, $B$ and $C$. Therefore, these matrices can be estimated with a Linear-Least Squares (LLS) approach. The LLS problem is then given by,
\begin{equation}
y^i_f = \begin{bmatrix} A^i&  B^i&  C^i& \end{bmatrix} \begin{bmatrix} y^i_p\\  u^i_p\\  u^i_f \end{bmatrix} = \Theta^i \phi^i
\end{equation}
Here $\Theta^i$ is the concatenation of all model matrices, and $\phi^i$ is the concatenation of $\yp$, $\uf$ and $\up$. We have chosen to use recursive LLS (RLS) to learn the model because we want to learn the system dynamics online and have the ability to track changes. The mathematical details of the RLS implementation can be found in \cite{haffert2021data}. With the models in hand, the controller has to be found. The cost function for the controller is the quadratic sum of all future measurements and control commands,

\begin{equation}
    J_i = y_f^{iT}y_f^{i} + \lambda u_f^{iT}u_f^{i} = \begin{bmatrix}
y_p^{iT} & u_p^{iT} & u_f^{iT}
\end{bmatrix} \begin{bmatrix}
A^TA & A^TB & A^TC \\ 
B^TA & B^TB & B^TC\\ 
C^TA & C^TB & C^TC
\end{bmatrix} 
\begin{bmatrix}
y_p^i \\
u_p^i \\
u_f^i
\end{bmatrix} 
+ \lambda u_f^{iT} u_f^i.
\end{equation}
Here $J_i$ is the cost at iteration $i$ and $\lambda$ is a regularization parameter that determines how much the DM commands have to be dampened. After some algebra we find the optimal control signal as,
\begin{equation}
  u_f = -\left(C^{iT} C^i + \lambda I\right)^{-1}\begin{bmatrix}
A^{iT}C & B^{iT}C^i
\end{bmatrix}
\begin{bmatrix}
y_p^i \\
u_p^i
\end{bmatrix} = -K^i \begin{bmatrix}
y_p^i \\
u_p^i
\end{bmatrix}.
\end{equation}
Here $K_i$ is the controller at time step $i$. The derivation that was presented here assumes that all measurements and commands are for a single actuator or mode. This assumes that there is no cross-coupling between the modes. This allows us to create a distributed controller. First the wavefront sensor reconstructs the modal coefficients, generally with an interaction matrix, and the DDSPC filters are applied to each mode independently. This decoupling makes this algorithm naturally parallel. However, there is no reason why a model with all modal coupling added back would not work. The same equations will govern this model. The only downside is the increased computational complexity.

The system is first trained before the algorithm is deployed on realistic disturbance. This is necessary to make sure the algorithm will not get stuck in Null spaces of the model. A System identification (SI) approach is used to let the algorithm get familiar with the system it is controlling. Here we excite the system with a known disturbance by adding noise to the controller commands. For high-order systems, it is important to construct information-rich signals that are able to persistently excite all relevant frequencies. A popular choice is a random binary signal (RBS). During the learning sequence, a randomly generated binary signal will be added to the final computed commands.

\subsection{Optical gain tracking}
One of the exciting aspects of online optimization of the controller is its ability to track changes in the system. These changes could be either non-stationary turbulence or changes in the instrument itself. The pyramid wavefront sensor \cite{ragazzoni1996pupil} is a non-linear wavefront sensor. One of the associated problems is a reduction of the optical gain when high-order turbulence is present. The relatively small dynamic range of the PWFS saturates due to the high-order uncontrolled wavefront aberrations, which reduced the sensitivity of the low-order controlled modes. The reduction of the optical gain is essentially an aspect of non-linear interaction. There are several approaches that are currently proposed to measure the optical gain and then modify the gain of the controller. The measurement processes are complicated or rely on accurate knowledge of turbulence statistics \cite{deo2019telescope, chambouleyron2020pyramid}. These may not always be available. The DDSPC controller can modify its gain during closed-loop operation, and would be a simple solution for the optical gain problem.

\begin{figure}
    \begin{center}
        \includegraphics{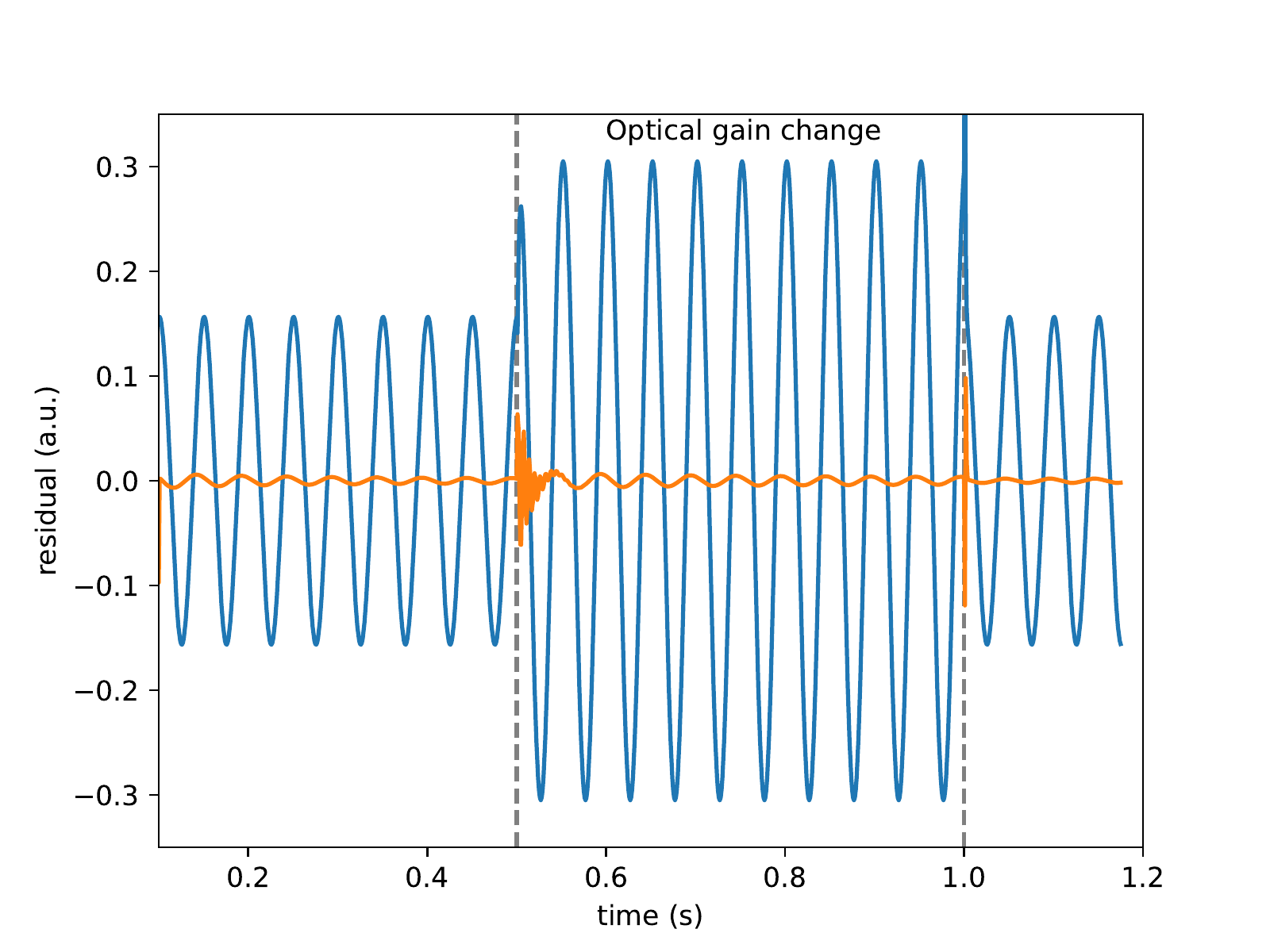}%
	\end{center}
    \caption{The residuals of a single-input-single-output system. The disturbance is a sinusoidal input with a frequency of 20 Hz. The integrator (in blue) had its gain optimized for this particular disturbance. At 0.5 s the relative optical gain of the sensor is changed to 0.5. And at 1.0 s the gain is set back to 1.0. The predictive controller is able to find the optimal feedback signal within several tens of ms during the first change. At the second optical gain change, the controller finds the optimal gain within several iterations. }
   \label{fig:optical_gain} 
\end{figure}

A toy model with a single controlled mode is shown in Figure \ref{fig:optical_gain}. The optical gain, or the sensor gain, is abruptly changed at 0.5 s. At 1.0 s the optical gain is set back again to its initial value. The predictive controller is able to find the optimal feedback signal within several tens of ms during the first change. At the second optical gain change, the controller finds the optimal gain within several iterations. This shows the power of online-learning and tracking of the system, changes in the system and the disturbance are pickup smoothly by the controller.

\subsection{Closing the loop on all modes}
\begin{figure}
    \begin{center}
        \includegraphics[width=\textwidth]{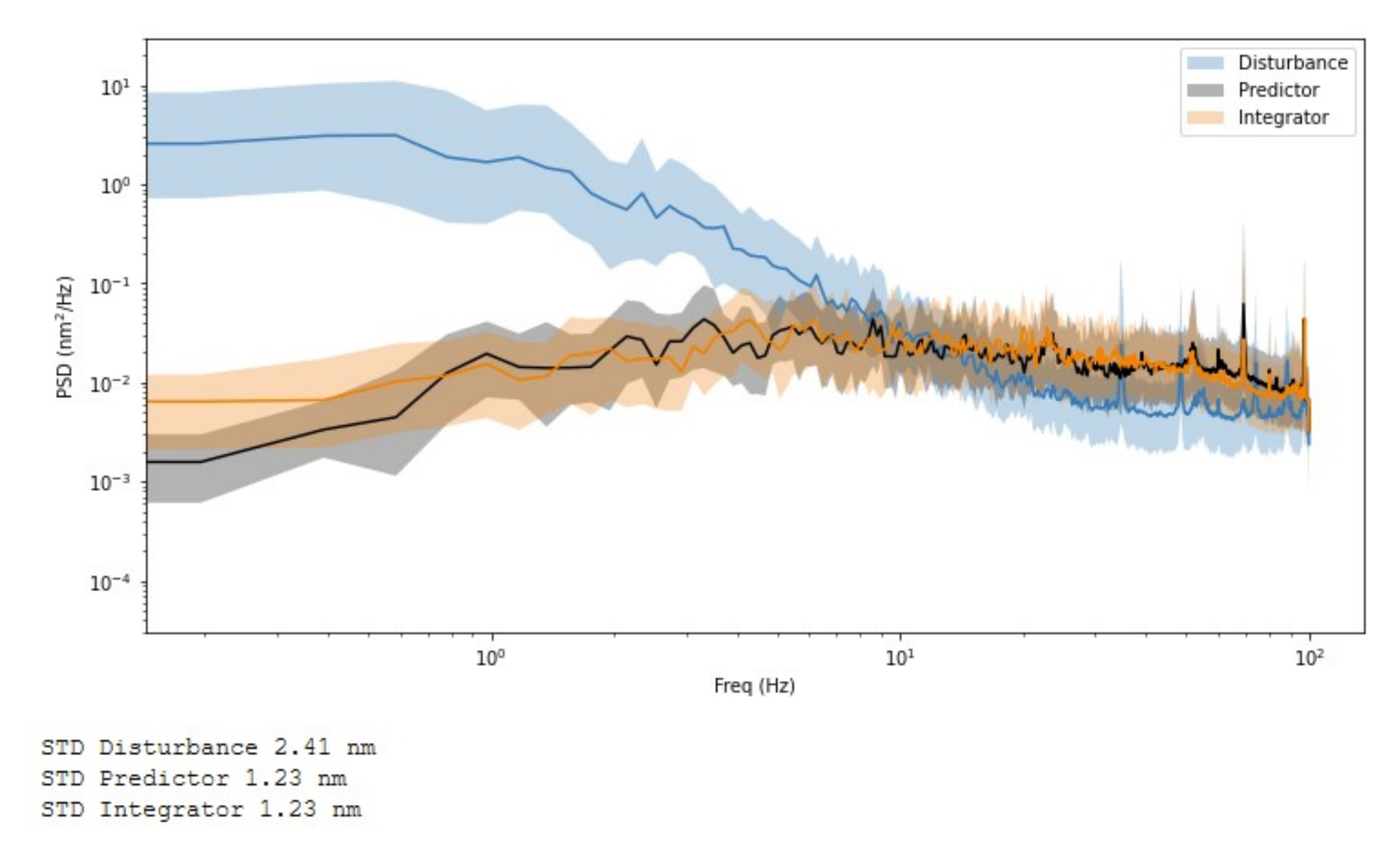}%
	\end{center}
    \caption{The power spectral density of the bench turbulence for MagAO-X before and after control. The solid line shows the median performance across all modes and the shaded area show the 16\% and 84\% confidence intervals. The predictive control approach reaches the same performance as the optimized integrator. Both controllers have a nearly flat residual PSD, indicating that most of the residuals are pure noise.}
   \label{fig:clc} 
\end{figure}

The controller is now able to run closed-loop on all the DM modes of MagAO-X on the internal source at high speed. The Power Spectral Density (PSD) before and after control are shown in Figure \ref{fig:clc}. The predictive controller reaches the same rms as the optimized controller. This can be explained by the fact that the integrator already reduces the PSD to a nearly flat PSD, which means that the residuals are pure noise. Changing the control would lead to no additional gain.

\section{Focal plane wavefront control with MagAO-X}
We have implemented a new approach to focal plane wavefront control for MagAO-X. This is based on pair-wise probing and Electric Field Conjugation (EFC) \cite{give2007broadband}. Pair-wise probing with the DM leads to a linear response between the probed images and the electric field:
\begin{equation}
    \Delta I = 4M\begin{bmatrix}
\Re{\left\{E\right\}}\\ 
\Im{\left\{E\right\}}
\end{bmatrix}.
\end{equation}
This system of equations can be inverted as long as there are enough probed images. The full electric field can be reconstructed from this. The EFC controller then tries to cancel the reconstructed electric field by injecting the opposite electric field with the DM. This problem uses the fact that there is also a linear relation between the DM modes and the electric field that the modes can create,
\begin{equation}
    \Delta E = G\vec{\alpha}.
\end{equation}
Here $G$ is the transfer matrix that transforms the modal coefficients $\alpha$ into the created electric field $\Delta E$. These two systems of linear equation are solved in succession to remove speckles. The main challenge for this approach is that both the system matrix $M$ and $G$ have to be modelled. That means that any error in the optical model will fold into a wrong reconstruction and control, which will limit the achieved contrast \cite{potier2020comparing}. However, both equations can also be combined into a single linear problem!
\begin{equation}
    \Delta I = 4MTG\alpha=F\alpha.
\end{equation}
Here $T$ is the matrix that separates an electric field into its real and imaginary components. We can now do an empirical calibration because there is a linear response between measurements and the DM. This means we don't have to rely on an optical model anymore.

\begin{figure}
    \begin{center}
        \includegraphics{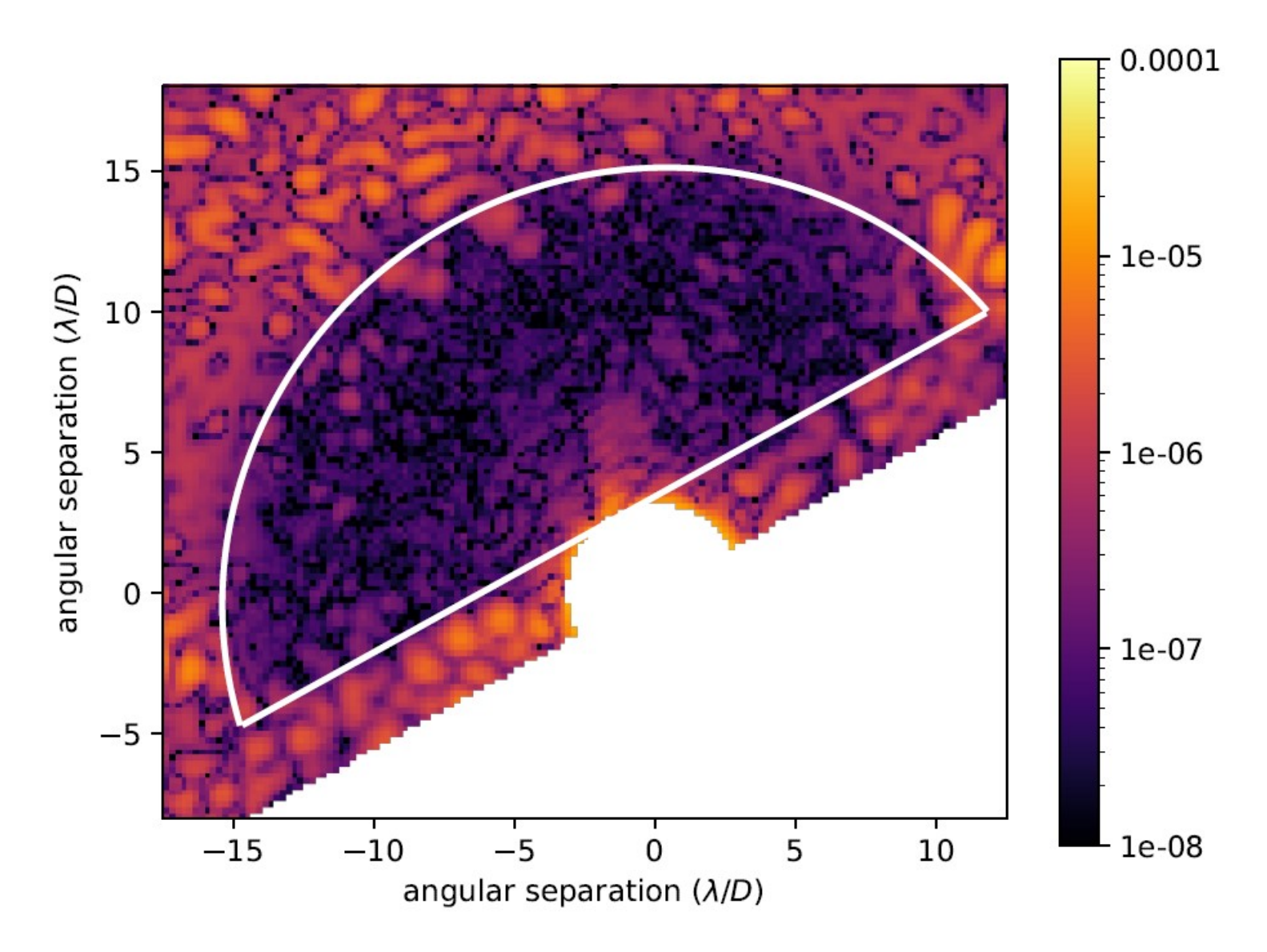}%
	\end{center}
    \caption{A dark hole created with the iEFC method and the PAPLC corongraph on MagAO-X. The dark hole is a D-shaped dark hole that runs from 3 $\lambda/D$ to 15 $\lambda/D$. The median contrast is $5\cdot 10^{-8}$. This was created in the H$\alpha$ narrow bandwidth filter.}
   \label{fig:darkhole} 
\end{figure}

The calibration of the interaction matrix, $F$, can be done by applying the classic push-and-pull method. We apply a positive amplitude for a mode that needs to be controlled. Then the pair-wise probed images are measured. This is then repeated with a negative amplitude mode. The two sets of pair-wise probed images can be subtracted from each other leading to the double difference images from which the columns of the interaction matrix are measured,
\begin{equation}
    F_i =  \frac{\Delta \Delta I}{2\alpha_i} =  \frac{\Delta I (+\alpha_i) - \Delta I (-\alpha_i)}{2\alpha_i}.
\end{equation}
The actual controller is then created by taking the pseudo-inverse of $F$. We call this approach implicit EFC (iEFC) because the electric field is not explicitly calculated anymore. Figure \ref{fig:darkhole} shows the deepest dark hole that we have made on MagAO-X. We have used a Phase Apodized Lyot Coronagraph \cite{por2020phase} in our H$\alpha$ filter to create the dark hole. The contrast is currently limited by tip/tilt jitter which is most likely created by our telescope simulator. We are currently exploring ways to remove these vibrations.

\section{Controlling piston with MagAO-X}
\begin{figure}
    \begin{center}
        \includegraphics[width=\textwidth]{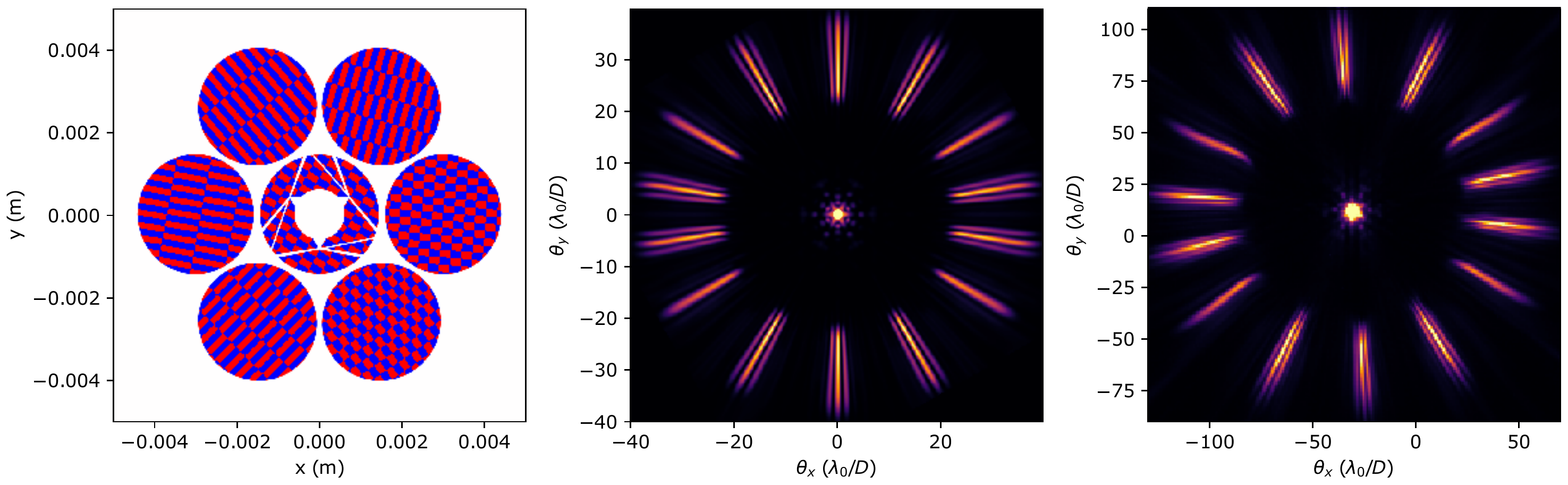}%
	\end{center}
    \caption{The designed phase pattern of the HDFS is shown on the left. The pattern has a $\pi$ phase shift between the red and blue colors. The model focal plane is shown in the middle panel and the right panel shows the measured focal plane of the fabricated HDFS.}
   \label{fig:hdfs} 
\end{figure}

The Holographic Dispersed Fringe Sensor (HDFS) is a new sensor that has been developped to measure differential piston on the GMT \cite{haffert2022phasing}. This sensor uses a hologram to combine pairs of segments of the GMT and it creates a dispersed fringe for each of the pairs. The piston can then be extracted from these dispersed fringes. An example of the HDFS is shown in Figure\ref{fig:hdfs}. The HDFS uses different multiplexed gratings on each segment to combine pairs and dispersed them at the same time. This is based on the concept of Holographic Aperture Masking \cite{doelman2021first}.

The HDFS was tested both in simulation and in the lab with MagAO-X where we reached less than 50 nm rms on the piston modes \cite{haffert2022phasing, hedglen2022lab}. It is now the 2nd stage wavefront sensor for the GMT facility AO system \cite{pacheco2022gmt}.

\section{Outlook and conclusion}
In this proceeding we describe several different parts of the wavefront control equation that we need to be able to handle if we want to search for Earth-like planets. We have implemented these approaches independently from each other in the lab. Each method will now be tested on-sky separately and finally combined at the same time. Combining both predictive control and focal plane wavefront control should lead to significantly improved sensitivity. We expect that we will be able to increase the sensitivity by 1 or 2 orders of magnitude after post-processing. We are planning an upgrade of the MagAO-X system in our next phase where each method will be implemented. We are also designing a high performance PIAACMC coronagraph that will allow use to do direct imaging at $1\lambda/D$. The combination of advanced wavefront control and a small-inner working angle coronagraph should help us to target Proxima b \cite{males2022magaox}. The results of these experiments will then be implemented into the design for GMagAO-X, which is a concept direct imaging instrument for the GMT\cite{males2022gmagaox, close2022gmagaoxDM, haffert2022gmagaox,kautz2022gmagaox}.

\acknowledgments 
Support for this work was provided by NASA through the NASA Hubble Fellowship grant \#HST-HF2-51436.001-A awarded by the Space Telescope Science Institute, which is operated by the Association of Universities for Research in Astronomy, Incorporated, under NASA contract NAS5-26555. This research made use of HCIPy, an open-source object-oriented framework written in Python for performing end-to-end simulations of high-contrast imaging instruments \cite{por2018high}. Support for this work was provided through the NSF "Cooperative Support Award \#2013059" and the AURA subaward NE0651C to GMTO.

\bibliography{report} 

\begin{thebibliography}{10}

\bibitem{traub2010exoplanets}
{Traub}, W.~A. and {Oppenheimer}, B.~R.,  [{\em {Direct Imaging of
  Exoplanets}}{\nolinebreak\hspace{0.1em}]},  111--156, {University of Arizona
  Press} (2010).

\bibitem{guyon2018extreme}
Guyon, O., ``Extreme adaptive optics,'' {\em Annual Review of Astronomy and
  Astrophysics}~{\bf 56},  315--355 (2018).

\bibitem{guyon2018wavefront}
Guyon, O., Mazin, B., Fitzgerald, M., Mawet, D., Marois, C., Skemer, A., Lozi,
  J., and Males, J., ``Wavefront control architecture and expected performance
  for the tmt planetary systems imager,'' in [{\em Adaptive Optics Systems
  VI}{\nolinebreak\hspace{0.1em}]},   {\bf 10703},  327--337, SPIE (2018).

\bibitem{beuzit2019sphere}
{Beuzit}, J.~L., {Vigan}, A., {Mouillet}, D., {Dohlen}, K., {Gratton}, R.,
  {Boccaletti}, A., {Sauvage}, J.~F., {Schmid}, H.~M., {Langlois}, M., {Petit},
  C., {Baruffolo}, A., {Feldt}, M., {Milli}, J., {Wahhaj}, Z., {Abe}, L.,
  {Anselmi}, U., {Antichi}, J., {Barette}, R., {Baudrand}, J., {Baudoz}, P.,
  {Bazzon}, A., {Bernardi}, P., {Blanchard}, P., {Brast}, R., {Bruno}, P.,
  {Buey}, T., {Carbillet}, M., {Carle}, M., {Cascone}, E., {Chapron}, F.,
  {Charton}, J., {Chauvin}, G., {Claudi}, R., {Costille}, A., {De Caprio}, V.,
  {de Boer}, J., {Delboulb{\'e}}, A., {Desidera}, S., {Dominik}, C., {Downing},
  M., {Dupuis}, O., {Fabron}, C., {Fantinel}, D., {Farisato}, G., {Feautrier},
  P., {Fedrigo}, E., {Fusco}, T., {Gigan}, P., {Ginski}, C., {Girard}, J.,
  {Giro}, E., {Gisler}, D., {Gluck}, L., {Gry}, C., {Henning}, T., {Hubin}, N.,
  {Hugot}, E., {Incorvaia}, S., {Jaquet}, M., {Kasper}, M., {Lagadec}, E.,
  {Lagrange}, A.~M., {Le Coroller}, H., {Le Mignant}, D., {Le Ruyet}, B.,
  {Lessio}, G., {Lizon}, J.~L., {Llored}, M., {Lundin}, L., {Madec}, F.,
  {Magnard}, Y., {Marteaud}, M., {Martinez}, P., {Maurel}, D., {M{\'e}nard},
  F., {Mesa}, D., {M{\"o}ller-Nilsson}, O., {Moulin}, T., {Moutou}, C.,
  {Orign{\'e}}, A., {Parisot}, J., {Pavlov}, A., {Perret}, D., {Pragt}, J.,
  {Puget}, P., {Rabou}, P., {Ramos}, J., {Reess}, J.~M., {Rigal}, F., {Rochat},
  S., {Roelfsema}, R., {Rousset}, G., {Roux}, A., {Saisse}, M., {Salasnich},
  B., {Santambrogio}, E., {Scuderi}, S., {Segransan}, D., {Sevin}, A.,
  {Siebenmorgen}, R., {Soenke}, C., {Stadler}, E., {Suarez}, M., {Tiph{\`e}ne},
  D., {Turatto}, M., {Udry}, S., {Vakili}, F., {Waters}, L.~B.~F.~M., {Weber},
  L., {Wildi}, F., {Zins}, G., and {Zurlo}, A., ``{SPHERE: the exoplanet imager
  for the Very Large Telescope},'' {\em \aap}~{\bf 631},  A155 (Nov 2019).

\bibitem{marley2007hotjupiters}
{Marley}, M.~S., {Fortney}, J.~J., {Hubickyj}, O., {Bodenheimer}, P., and
  {Lissauer}, J.~J., ``{On the Luminosity of Young Jupiters},'' {\em \apj}~{\bf
  655},  541--549 (Jan. 2007).

\bibitem{bowler2015gpoccurence}
{Bowler}, B.~P., {Liu}, M.~C., {Shkolnik}, E.~L., and {Tamura}, M., ``{Planets
  around Low-mass Stars (PALMS). IV. The Outer Architecture of M Dwarf
  Planetary Systems},'' {\em \apjs}~{\bf 216},  7 (Jan 2015).

\bibitem{nielsen2019gpies}
{Nielsen}, E.~L., {De Rosa}, R.~J., {Macintosh}, B., {Wang}, J.~J., {Ruffio},
  J.-B., {Chiang}, E., {Marley}, M.~S., {Saumon}, D., {Savransky}, D.,
  {Ammons}, S.~M., {Bailey}, V.~P., {Barman}, T., {Blain}, C., {Bulger}, J.,
  {Burrows}, A., {Chilcote}, J., {Cotten}, T., {Czekala}, I., {Doyon}, R.,
  {Duch{\^e}ne}, G., {Esposito}, T.~M., {Fabrycky}, D., {Fitzgerald}, M.~P.,
  {Follette}, K.~B., {Fortney}, J.~J., {Gerard}, B.~L., {Goodsell}, S.~J.,
  {Graham}, J.~R., {Greenbaum}, A.~Z., {Hibon}, P., {Hinkley}, S., {Hirsch},
  L.~A., {Hom}, J., {Hung}, L.-W., {Dawson}, R.~I., {Ingraham}, P., {Kalas},
  P., {Konopacky}, Q., {Larkin}, J.~E., {Lee}, E.~J., {Lin}, J.~W., {Maire},
  J., {Marchis}, F., {Marois}, C., {Metchev}, S., {Millar-Blanchaer}, M.~A.,
  {Morzinski}, K.~M., {Oppenheimer}, R., {Palmer}, D., {Patience}, J.,
  {Perrin}, M., {Poyneer}, L., {Pueyo}, L., {Rafikov}, R.~R., {Rajan}, A.,
  {Rameau}, J., {Rantakyr{\"o}}, F.~T., {Ren}, B., {Schneider}, A.~C.,
  {Sivaramakrishnan}, A., {Song}, I., {Soummer}, R., {Tallis}, M., {Thomas},
  S., {Ward-Duong}, K., and {Wolff}, S., ``{The Gemini Planet Imager Exoplanet
  Survey: Giant Planet and Brown Dwarf Demographics from 10 to 100 au},'' {\em
  \aj}~{\bf 158},  13 (July 2019).

\bibitem{fernandes2019occurence}
{Fernandes}, R.~B., {Mulders}, G.~D., {Pascucci}, I., {Mordasini}, C., and
  {Emsenhuber}, A., ``{Hints for a Turnover at the Snow Line in the Giant
  Planet Occurrence Rate},'' {\em \apj}~{\bf 874},  81 (Mar. 2019).

\bibitem{wagner2019wideoccurence}
{Wagner}, K., {Apai}, D., and {Kratter}, K.~M., ``{On the Mass Function,
  Multiplicity, and Origins of Wide-orbit Giant Planets},'' {\em \apj}~{\bf
  877},  46 (May 2019).

\bibitem{kasper2012hci}
{Kasper}, M., ``{Adaptive optics for high contrast imaging},'' in [{\em
  \procspie}{\nolinebreak\hspace{0.1em}]},  {\em Society of Photo-Optical
  Instrumentation Engineers (SPIE) Conference Series} {\bf 8447},  84470B
  (2012).

\bibitem{milli2017sphereperformance}
{Milli}, J., {Mouillet}, D., {Fusco}, T., {Girard}, J.~H., {Masciadri}, E.,
  {Pena}, E., {Sauvage}, J.~F., {Reyes}, C., {Dohlen}, K., {Beuzit}, J.~L.,
  {Kasper}, M., {Sarazin}, M., and {Cantalloube}, F., ``{Performance of the
  extreme-AO instrument VLT/SPHERE and dependence on the atmospheric
  conditions},'' {\em arXiv e-prints} ,  arXiv:1710.05417 (Oct. 2017).

\bibitem{cantalloube2019winddrivenhalo}
{Cantalloube}, F., {Absil}, O., {Bertram}, T., {Brandner}, W., {Delacroix}, C.,
  {Feldt}, M., {Kenworthy}, M., {Kulas}, M., {Milli}, J., {Neureuther}, P.,
  {Orban de Xivry}, G., {Pathak}, P., {Por}, E., {Scheithauer}, S., {Steuer},
  H., and {van Boekel}, R., ``{High contrast imaging with ELT/METIS: The wind
  driven halo, from SPHERE to METIS},'' {\em arXiv e-prints} ,
  arXiv:1911.11241 (Nov. 2019).

\bibitem{males2022magaox}
Males, J.~R., Close, L.~M., Haffert, S.~Y., Long, J., Schatz, L., Gorkom,
  K.~V., Guyon, O., Hedglen, A.~D., Kautz, M., Pearce, L., Trzaska, J.,
  Lumbres, J., Rodack, A., Kueny, J., and Foster, W., ``{MagAO-X: current
  status and plans for Phase II},'' in [{\em Adaptive Optics Systems
  VIII}{\nolinebreak\hspace{0.1em}]},   {\bf 12185}, International Society for
  Optics and Photonics, SPIE (2022).

\bibitem{jovanovic2015scexao}
{Jovanovic}, N., {Martinache}, F., {Guyon}, O., {Clergeon}, C., {Singh}, G.,
  {Kudo}, T., {Garrel}, V., {Newman}, K., {Doughty}, D., {Lozi}, J., {Males},
  J., {Minowa}, Y., {Hayano}, Y., {Takato}, N., {Morino}, J., {Kuhn}, J.,
  {Serabyn}, E., {Norris}, B., {Tuthill}, P., {Schworer}, G., {Stewart}, P.,
  {Close}, L., {Huby}, E., {Perrin}, G., {Lacour}, S., {Gauchet}, L.,
  {Vievard}, S., {Murakami}, N., {Oshiyama}, F., {Baba}, N., {Matsuo}, T.,
  {Nishikawa}, J., {Tamura}, M., {Lai}, O., {Marchis}, F., {Duchene}, G.,
  {Kotani}, T., and {Woillez}, J., ``{The Subaru Coronagraphic Extreme Adaptive
  Optics System: Enabling High-Contrast Imaging on Solar-System Scales},'' {\em
  \pasp}~{\bf 127},  890 (Sep 2015).

\bibitem{males2020magao}
Males, J.~R., Close, L.~M., Guyon, O., Hedglen, A.~D., Van~Gorkom, K., Long,
  J.~D., Kautz, M., Lumbres, J., Schatz, L., Rodack, A., et~al., ``Magao-x
  first light,'' in [{\em Adaptive Optics Systems
  VII}{\nolinebreak\hspace{0.1em}]},   {\bf 11448},  114484L, International
  Society for Optics and Photonics (2020).

\bibitem{guyon2017eof}
{Guyon}, O. and {Males}, J., ``{Adaptive Optics Predictive Control with
  Empirical Orthogonal Functions (EOFs)},'' {\em arXiv e-prints} ,
  arXiv:1707.00570 (July 2017).

\bibitem{males2018lpc}
{Males}, J.~R. and {Guyon}, O., ``{Ground-based adaptive optics coronagraphic
  performance under closed-loop predictive control},'' {\em Journal of
  Astronomical Telescopes, Instruments, and Systems}~{\bf 4},  019001 (Jan.
  2018).

\bibitem{correia2020hcipwfs}
{Correia}, C.~M., {Fauvarque}, O., {Bond}, C.~Z., {Chambouleyron}, V.,
  {Sauvage}, J.-F., and {Fusco}, T., ``{Performance limits of
  adaptive-optics/high-contrast imagers with pyramid wavefront sensors},'' {\em
  \mnras}~{\bf 495},  4380--4391 (June 2020).

\bibitem{haffert2021data}
Haffert, S.~Y., Males, J.~R., Close, L.~M., Van~Gorkom, K., Long, J.~D.,
  Hedglen, A.~D., Guyon, O., Schatz, L., Kautz, M.~Y., Lumbres, J., et~al.,
  ``Data-driven subspace predictive control of adaptive optics for
  high-contrast imaging,'' {\em Journal of Astronomical Telescopes,
  Instruments, and Systems}~{\bf 7}(2),  029001 (2021).

\bibitem{cuda}
nVIDIA, ``{CUDA Toolkit}.'' \url{https://developer.nvidia.com/cuda-toolkit}
  (2021).
\newblock [Online; accessed 9-August-2021].

\bibitem{ragazzoni1996pupil}
Ragazzoni, R., ``Pupil plane wavefront sensing with an oscillating prism,''
  {\em Journal of modern optics}~{\bf 43}(2),  289--293 (1996).

\bibitem{deo2019telescope}
Deo, V., Gendron, {\'E}., Rousset, G., Vidal, F., Sevin, A., Ferreira, F.,
  Gratadour, D., and Buey, T., ``A telescope-ready approach for modal
  compensation of pyramid wavefront sensor optical gain,'' {\em Astronomy \&
  Astrophysics}~{\bf 629},  A107 (2019).

\bibitem{chambouleyron2020pyramid}
Chambouleyron, V., Fauvarque, O., Janin-Potiron, P., Correia, C., Sauvage,
  J.-F., Schwartz, N., Neichel, B., and Fusco, T., ``Pyramid wavefront sensor
  optical gains compensation using a convolutional model,'' {\em Astronomy \&
  Astrophysics}~{\bf 644},  A6 (2020).

\bibitem{give2007broadband}
Give'on, A., Kern, B., Shaklan, S., Moody, D.~C., and Pueyo, L., ``Broadband
  wavefront correction algorithm for high-contrast imaging systems,'' in [{\em
  Astronomical Adaptive Optics Systems and Applications
  III}{\nolinebreak\hspace{0.1em}]},   {\bf 6691},  63--73, SPIE (2007).

\bibitem{potier2020comparing}
Potier, A., Baudoz, P., Galicher, R., Singh, G., and Boccaletti, A.,
  ``Comparing focal plane wavefront control techniques: Numerical simulations
  and laboratory experiments,'' {\em Astronomy \& Astrophysics}~{\bf 635},
  A192 (2020).

\bibitem{por2020phase}
Por, E.~H., ``Phase-apodized-pupil lyot coronagraphs for arbitrary telescope
  pupils,'' {\em The Astrophysical Journal}~{\bf 888}(2),  127 (2020).

\bibitem{haffert2022phasing}
Haffert, S.~Y., Close, L.~M., Hedglen, A.~D., Males, J.~R., Kautz, M., Bouchez,
  A.~H., Demers, R., Quir{\'o}s-Pacheco, F., Sitarski, B.~N., Van~Gorkom, K.,
  et~al., ``Phasing the giant magellan telescope with the holographic dispersed
  fringe sensor,'' {\em Journal of Astronomical Telescopes, Instruments, and
  Systems}~{\bf 8}(2),  021513 (2022).

\bibitem{doelman2021first}
Doelman, D.~S., Wardenier, J.~P., Tuthill, P., Fitzgerald, M.~P., Lyke, J.,
  Sallum, S., Norris, B., Warriner, N.~Z., Keller, C., Escuti, M.~J., et~al.,
  ``First light of a holographic aperture mask: Observation at the keck osiris
  imager,'' {\em Astronomy \& Astrophysics}~{\bf 649},  A168 (2021).

\bibitem{hedglen2022lab}
Hedglen, A.~D., Close, L.~M., Haffert, S.~Y., Males, J.~R., Kautz, M., Bouchez,
  A.~H., Demers, R., Quir{\'o}s-Pacheco, F., Sitarski, B.~N., Guyon, O.,
  et~al., ``Lab tests of segment/petal phasing with a pyramid wavefront sensor
  and a holographic dispersed fringe sensor in turbulence with the giant
  magellan telescope high contrast adaptive optics phasing testbed,'' {\em
  Journal of Astronomical Telescopes, Instruments, and Systems}~{\bf 8}(2),
  021515 (2022).

\bibitem{pacheco2022gmt}
Quiros-Pacheco, F., van Dam, M., Bouchez, A.~H., Conan, R., Haffert, S.~Y.,
  Agapito, G., , and Demers, R., ``The giant magellan telescope natural
  guidestar adaptive optics mode: improving the robustness of segment piston
  control,'' in [{\em Adaptive Optics Systems
  VIII}{\nolinebreak\hspace{0.1em}]},   {\bf 12185}, International Society for
  Optics and Photonics, SPIE (2022).

\bibitem{males2022gmagaox}
Males, J.~R., Close, L.~M., Haffert, S.~Y., Guyon, O., Coronado, F., Gasho, V.,
  Noenickx, J., Ford, J., Kelly, D., and Durney, O., ``{The conceptual design
  of GMagAO-X: visible wavelength high contrast imaging with GMT},'' in [{\em
  Adaptive Optics Systems VIII}{\nolinebreak\hspace{0.1em}]},   {\bf 12185},
  International Society for Optics and Photonics, SPIE (2022).

\bibitem{close2022gmagaoxDM}
Close, L.~M., Males, J.~R., Durney, O., Coronado, F., Haffert, S.~Y., Gasho,
  V., Hedglen, A., Kautz, M., Connors, T., Sullivan, M., Guyon, O., and
  Noenickx, J., ``{The Optical and Mechanical Design for the 21,000 Actuator
  ExAO System for the Giant Magellan Telescope: GMagAO-X},'' in [{\em Adaptive
  Optics Systems VIII}{\nolinebreak\hspace{0.1em}]},   {\bf 12185},
  International Society for Optics and Photonics, SPIE (2022).

\bibitem{haffert2022gmagaox}
Haffert, S.~Y., Males, J.~R., Close, L.~M., Guyon, O., Hedglen, A., and Kautz,
  M., ``{Visible extreme adaptive optics for GMagAO-X with the triple-stage AO
  architecture (TSAO)},'' in [{\em Adaptive Optics Systems
  VIII}{\nolinebreak\hspace{0.1em}]},   {\bf 12185}, International Society for
  Optics and Photonics, SPIE (2022).

\bibitem{kautz2022gmagaox}
Kautz, M., Close, L.~M., Hedglen, A., Haffert, S.~Y., Males, J.~R., and
  Coronado, F., ``{A Novel Hexpyramid Pupil Slicer for an ExAO Parallel DM for
  the Giant Magellan Telescope},'' in [{\em Adaptive Optics Systems
  VIII}{\nolinebreak\hspace{0.1em}]},   {\bf 12185}, International Society for
  Optics and Photonics, SPIE (2022).

\bibitem{por2018high}
Por, E.~H., Haffert, S.~Y., Radhakrishnan, V.~M., Doelman, D.~S., van Kooten,
  M., and Bos, S.~P., ``High contrast imaging for python (hcipy): an
  open-source adaptive optics and coronagraph simulator,'' in [{\em Adaptive
  Optics Systems VI}{\nolinebreak\hspace{0.1em}]},   {\bf 10703},  1070342,
  International Society for Optics and Photonics (2018).

\end{thebibliography}
\bibliographystyle{spiebib} 

\end{document}